\documentclass[amsmath,aps,showpacs,a4paper,10pt]{revtex4}

 \usepackage{epsf}
 \usepackage{graphicx}    

 \usepackage{enumerate}   

\begin{document}

 \newcommand{\be}[1]{\begin{equation}\label{#1}}
 \newcommand{\ee}{\end{equation}}
 \newcommand{\bea}{\begin{eqnarray}}
 \newcommand{\eea}{\end{eqnarray}}
 \def\disp{\displaystyle}

 \def\gsim{ \lower .75ex \hbox{$\sim$} \llap{\raise .27ex \hbox{$>$}} }
 \def\lsim{ \lower .75ex \hbox{$\sim$} \llap{\raise .27ex \hbox{$<$}} }

 \begin{titlepage}

 \begin{flushright}
 arXiv:1902.03580
 \end{flushright}

 \title{\Large \bf Cosmic Anisotropy and Fast Radio Bursts}

 \author{Da-Chun~Qiang\,}
 \email[\,email address:\ ]{875019424@qq.com}
 \affiliation{School of Physics,
 Beijing Institute of Technology, Beijing 100081, China}

 \author{Hua-Kai~Deng\,}
 \email[\,email address:\ ]{dhklook@163.com}
 \affiliation{School of Physics,
 Beijing Institute of Technology, Beijing 100081, China}

 \author{Hao~Wei\,}
 \email[\,Corresponding author;\ email address:\ ]{haowei@bit.edu.cn}
 \affiliation{School of Physics,
 Beijing Institute of Technology, Beijing 100081, China}

 \begin{abstract}\vspace{1cm}
 \centerline{\bf ABSTRACT}\vspace{2mm}
 In the recent years, the field of fast radio bursts (FRBs) is
 thriving and growing rapidly. It is of interest to study cosmology
 by using FRBs with known redshifts. In the present work, we try to
 test the possible cosmic anisotropy with the simulated FRBs.
 In particular, we only consider the possible dipole in FRBs,
 rather than the cosmic anisotropy in general, while the analysis is
 only concerned with finding the rough number of necessary data points
 to distinguish a dipole from a monopole structure through simulations.
 Noting that there is no a large sample of actual data of FRBs with
 known redshifts by now, simulations are necessary to this end.
 We find that at least 2800, 190, 100 FRBs are competent to find the
 cosmic dipole with amplitude 0.01, 0.03, 0.05, respectively.
 Unfortunately, even 10000 FRBs are not competent to find the
 tiny cosmic dipole with amplitude of ${\cal O}(10^{-3})$. On the other
 hand, at least 20 FRBs with known redshifts are competent to find the
 cosmic dipole with amplitude 0.1. We expect that such a big cosmic
 dipole could be ruled out by using only a few tens of FRBs
 with known redshifts in the near future.
 \end{abstract}


 \maketitle

 \end{titlepage}

 \renewcommand{\baselinestretch}{1.0}


\section{Introduction}\label{sec1}

In the past few years, the newly discovered fast radio bursts (FRBs)
 have become a promising field in astronomy and cosmology, which is
 currently thriving and growing rapidly~\cite{Keane:2018jqo,
 Lorimer:2018rwi,Pen:2018ilo,Kulkarni:2018ola,Burke-Spolaor:2018xoa,
 Macquart:2018fhn,NAFRBs}. In fact, since its
 first discovery~\cite{Lorimer:2007qn}, more and more evidences
 suggest that FRBs are at cosmological distances
 (see e.g.~\cite{Thornton:2013iua,Champion:2015pmj,Spitler:2016dmz,
 Tendulkar:2017vuq,Marcote:2017wan,Chatterjee:2017dqg,
 Keane:2016yyk,Shannon:2018ASKAP,Amiri:2019qbv,Amiri:2019bjk,
 Petroff:2016tcr}). So, it is reasonable to consider the cosmological
 application of FRBs.

One of the key measured quantities of FRBs is the so-called dispersion
 measure (DM). According to the textbook~\cite{Rybicki:1979} (see also
 e.g.~\cite{Deng:2013aga,Yang:2016zbm,Ioka:2003fr,Inoue:2003ga}), an
 electromagnetic signal of frequency $\nu$ propagates through
 an ionized medium (plasma) with a velocity $\disp v=c\, (1-
 \nu_{p\; }^2/\nu^2)^{1/2}$, less than the speed of light
 in vacuum~$c$, where $\disp \nu_p=(n_e\, e^2 /\pi m_e)^{1/2}
 \simeq 8.98\times 10^3\, n_e^{1/2}\; {\rm Hz}$ is the plasma
 frequency, $n_e$ is the number density of free electrons in
 the medium (given in units of $\rm cm^{-3}$), $m_e$ and $e$
 are the mass and charge of electron, respectively. Therefore,
 an electromagnetic signal with frequency $\nu\gg\nu_p$ is
 delayed relative to a signal in vacuum by a
 time~\cite{Rybicki:1979,Deng:2013aga,Yang:2016zbm,Ioka:2003fr,Inoue:2003ga}
 \be{eq1}
 t_p\simeq\int\frac{\nu_p^2}{2\nu^2}\frac{dl}{c}=
 \frac{e^2}{2\pi m_{e\,}c}\frac{1}{\nu^2}\int n_e\,dl\equiv
 \frac{e^2}{2\pi m_{e\,}c}\frac{\rm DM}{\nu^2}
 \simeq 4.2\,{\rm s}\,\left(\frac{\nu}{\rm GHz}\right)^{-2}
 \frac{\rm DM}{10^3\;{\rm pc\hspace{0.24em} cm^{-3}}}\,,
 \ee
 where the dispersion measure ${\rm DM}=\int n_e\,dl$ means actually
 the column density of the free electrons. In practice, it is convenient
 to measure the time delay between two frequencies. For a plasma at
 redshift $z$, the rest frame (infinitesimal) time delay between two
 rest frame frequencies is given by
 \be{eq2}
 d t_z=\frac{e^2}{2\pi m_{e\,}c}\,\bigg(\frac{1}{\nu_{1,\,z}^2}
 -\frac{1}{\nu_{2,\,z}^2}\bigg)\,n_{e,\,z}\,dl\,.
 \ee
 In the observer frame, the observed time delay and the
 observed frequency are both redshifted, namely $d t=d t_z\,(1+z)$, and
 $\nu=\nu_z/(1+z)$. So, the the observed time delay is given
 by~\cite{Deng:2013aga,Yang:2016zbm,Ioka:2003fr,Inoue:2003ga}
 \be{eq3}
 \Delta t=\int dt=\frac{e^2}{2\pi m_{e\,}c}\left(
 \frac{1}{\nu_{1}^2}-\frac{1}{\nu_{2}^2}\right)
 \int\frac{n_{e,\,z}}{1+z}\,dl\equiv
 \frac{e^2}{2\pi m_{e\,}c}\left(
 \frac{1}{\nu_{1}^2}-\frac{1}{\nu_{2}^2}\right){\rm DM}\,,
 \ee
 and the observed dispersion
 measure reads~\cite{Deng:2013aga,Yang:2016zbm,Ioka:2003fr,Inoue:2003ga}
 \be{eq4}
 {\rm DM}=\int\frac{n_{e,\,z}}{1+z}\,dl\,.
 \ee
 Using Eq.~(\ref{eq3}), one can observationally obtain DM by
 measuring $\Delta t$ between two frequencies $\nu_1$ and
 $\nu_2$. On the other hand, since the distance $dl$ along
 the path in Eq.~(\ref{eq4}) records the expansion history of
 the universe (see below in details), the observed DM can
 be used to study cosmology. It is worth noting that DM is
 traditionally used to observe pulsars in or nearby our Galaxy
 (Milky Way) and other objects, but now is also
 extended to FRBs since the first day that FRB was discovered.

For typical FRBs, the observed DM is ${\cal O}(10^2)$ or
 ${\cal O}(10^3)$ $\rm pc\hspace{0.24em} cm^{-3}$
 with negligible uncertainties of ${\cal O}(10^{-1})$ or
 ${\cal O}(10^{-2})$ $\rm pc\hspace{0.24em} cm^{-3}$~\cite{Petroff:2016tcr},
 and $\nu\sim{\rm GHz}$, $\Delta t\sim {\rm ms}$. Most of the
 published FRBs are at high Galactic latitude
 $|b|>10^\circ$~\cite{Petroff:2016tcr}, which is helpful to
 minimize the contribution of our Galaxy (Milky Way) to the
 electron column density, DM. As of 1 January 2020, 107
 FRBs have been found~\cite{Petroff:2016tcr} mainly by
 the~telescopes Parkes, UTMOST, ASKAP and CHIME. In particular,
 the number of FRBs increased rapidly after the (pre-)commissions of
 ASKAP and CHIME in 2018. In fact, the lower-limit estimates
 for the number of the FRB events occurring are a few thousand
 each day~\cite{Keane:2018jqo,Bhandari:2017qrj}. Even conservatively,
 the FRB event rate floor derived from the pre-commissioning of
 CHIME is $3\times 10^2$ events per day~\cite{Amiri:2019qbv}.
 Thus, the observed FRBs will be numerous in the coming years.

As a very crude rule of thumb, the redshift $z\sim {\rm DM}/(1000\;
 \rm pc\hspace{0.24em} cm^{-3})$~\cite{Lorimer:2018rwi}.
 For all the 107 observed~FRBs to date, their DMs are in the
 range $100\sim 2600\;\rm pc\hspace{0.24em} cm^{-3}$
 approximately~\cite{Petroff:2016tcr}, and hence one can infer
 redshifts in the range $0.1\lesssim z\lesssim 2.6$ crudely.
 In fact, the redshift of FRB 121102~\cite{Spitler:2016dmz,
 Tendulkar:2017vuq,Marcote:2017wan,Chatterjee:2017dqg} has been
 directly identified ($z=0.19273$~\cite{Tendulkar:2017vuq}), which is
 the only repeating FRB source before CHIME. Recently, the other 9 repeating
 FRB sources have been found by CHIME~\cite{Amiri:2019bjk,Andersen:2019yex},
 and it is reasonable to expect more repeating FRB sources in the future. As
 mentioned in e.g.~\cite{Yang:2016zbm}, there exist three possibilities to
 identify FRB redshifts in the future, namely ({\it i}) pin down the precise
 location and then the possible host galaxy of FRB (especially repeating
 FRB) by using VLBI observations~\cite{Marcote:2019sjf}. ({\it ii}) catch
 the afterglow of FRB by performing multi-wavelength follow-up observations
 soon after the FRB trigger~\cite{Yi:2014jka}. ({\it iii}) detect the FRB
 counterparts in other wavelengths (for example, gamma-ray bursts (GRBs)).
 Since the field of FRBs is growing rapidly, similar to the history of
 GRBs~\cite{Kulkarni:2018ola}, numerous FRBs with identified redshifts might
 be available in the coming years.

It is worth noting that the up-to-date online catalogue of the
 observed FRBs can be found in~\cite{Petroff:2016tcr}, which
 summarizes almost all observational aspects concerning the
 published FRBs. On the other hand, in the literature, a lot of
 theoretical models have been proposed for FRBs, and the number
 of FRB theories is still increasing. Since in the present work
 we are mainly interested in the observational aspects, we just
 refer to~\cite{Platts:2018hiy} for the up-to-date online catalogue
 of FRB theories.

With the observed DMs and the identified redshifts, FRBs can be
 used to study cosmology. Actually, in the literature,
 there are some interesting works using FRBs in cosmology,
 e.g.~\cite{Deng:2013aga,Yang:2016zbm,Gao:2014iva,Zhou:2014yta,
 Yu:2017beg,Yang:2017bls,Wei:2018cgd,Li:2017mek,
 Jaroszynski:2018vgh,Madhavacheril:2019buy,Wang:2018ydd,
 Walters:2017afr,Cai:2019cfw}. In the present work, we are interested in
 using the simulated FRBs to test the cosmological principle,
 which is one of the pillars of modern cosmology.

As a fundamental assumption, although the cosmological principle is
 indeed a very good approximation across a vast part of the universe
 (see e.g.~\cite{Hogg:2004vw,Hajian:2006ud}), actually it has
 not yet been well proven on cosmic scales
 $\gtrsim 1\,$Gpc~\cite{Caldwell:2007yu}. Therefore, it is still of
 interest to test both the homogeneity and the isotropy of the
 universe carefully. In fact, they could be broken in some
 theoretical models, such as the well-known
 Lema\^{\i}tre-Tolman-Bondi~(LTB) void model~\cite{LTB} (see
 also e.g.~\cite{Yan:2014eca,Deng:2018yhb,Deng:2018jrp} and
 references therein) violating the cosmic homogeneity, and the
 exotic G\"odel universe~\cite{Godel:1949ga} (see also
 e.g.~\cite{Li:2016nnn} and references therein), most of
 Bianchi type I$\,\sim\,$IX universes~\cite{Bianchi}, Finsler
 universe~\cite{Li:2015uda}, violating the cosmic isotropy.
 On the other hand, many observational hints of the cosmic
 inhomogeneity and/or anisotropy have been claimed in the
 literature (see e.g.~\cite{Deng:2018yhb,Deng:2018jrp} for
 brief reviews).

Here, we mainly concentrate on the possible cosmic anisotropy.
 In the past 15 years, various hints for the cosmic anisotropy
 have been found, for example, it is claimed that there exists
 a preferred direction in the cosmic microwave background (CMB)
 temperature map (known as the ``\,Axis~of~Evil\,'' in the
 literature)~\cite{Axisofevil,Zhao:2016fas,Hansen:2004vq}, the
 distribution of type Ia supernovae
 (SNIa)~\cite{Schwarz:2007wf,Antoniou:2010gw,
 Mariano:2012wx,Cai:2011xs,Zhao:2013yaa,Yang:2013gea,Chang:2014nca,
 Lin:2015rza,Chang:2017bbi,Javanmardi:2015sfa,Lin:2016jqp,
 Bengaly:2015dza,Deng:2018yhb,Deng:2018jrp,Sun:2018cha,
 Andrade:2018eta,Ghodsi:2016dwp}, GRBs~\cite{Meszaros:2009ux,
 Wang:2014vqa,Chang:2014jza}, quasars and radio
 galaxies~\cite{Singal:2013aga,Bengaly:2017slg,Tiwari:papers,
 Singal:2011dy}, rotationally supported galaxies~\cite{Zhou:2017lwy,
 Chang:2018vxs}, and the quasar optical polarization data
 \cite{Hutsemekers,Pelgrims:2016mhx}. In addition, using the absorption
 systems in the spectra of distant quasars, it is claimed that the fine
 structure ``\,constant\,'' $\alpha$ is not only time-varying~\cite{Webb98,
 Webb00} (see also e.g.~\cite{{Uzan10,Barrow09,HWalpha}}), but also
 spatially varying~\cite{King:2012id,Webb:2010hc}. Precisely speaking,
 there also exists a preferred direction in the data of
 $\Delta\alpha/\alpha$. Interestingly, it is found in~\cite{Mariano:2012wx}
 that the preferred direction in $\Delta\alpha/\alpha$ might be correlated
 with the one in the distribution of SNIa.

Since the field of FRBs is growing rapidly today, and it can
 be used to study cosmology, in this work we try to test the
 possible cosmic anisotropy by using the simulated FRBs. In
 Sec.~\ref{sec2}, we briefly describe the methodology to
 simulate FRBs. In Sec.~\ref{sec3}, we test the cosmic
 anisotropy with the simulated FRBs. In Sec.~\ref{sec4}, some
 concluding remarks are given.


\section{Methodology to simulate FRBs}\label{sec2}

Clearly, the observed DM of FRB is given
 by~\cite{Deng:2013aga,Yang:2016zbm,Gao:2014iva,Zhou:2014yta,Yang:2017bls}
 \be{eq5}
 \rm DM_{obs}=DM_{MW}+DM_{IGM}+DM_{HG}\,,
 \ee
 where $\rm DM_{MW}$, $\rm DM_{IGM}$, $\rm DM_{HG}$ are the
 contributions from Milky Way, intergalactic medium (IGM),
 host galaxy (HG, actually including interstellar medium of
 HG and the near-source plasma) of FRB, respectively. In fact,
 $\rm DM_{MW}$ can be well constrained with the pulsar
 data~\cite{Taylor:1993my,Manchester:2004bp}. It strongly
 depends on Galactic latitude $|b|$, and has a maximum
 $\sim 10^3\;{\rm pc\hspace{0.24em} cm^{-3}}$ around $|b|\sim 0^\circ$,
 but becomes less than $100\;{\rm pc\hspace{0.24em} cm^{-3}}$ at high
 Galactic latitude $|b|>10^\circ$~\cite{Taylor:1993my,Manchester:2004bp}.
 As mentioned above, most of the published FRBs are at high
 Galactic latitude $|b|>10^\circ$~\cite{Petroff:2016tcr}, and hence
 $\rm DM_{MW}$ is a relatively small term in Eq.~(\ref{eq5}).
 For a well-localized FRB (available in the coming years), the
 corresponding $\rm DM_{MW}$ can be extracted with reasonable
 certainty~\cite{Cordes:2003ik,Cordes:2002wz,YMW16}. Thus, we
 can subtract this known $\rm DM_{MW}$ from $\rm DM_{obs}$. Following
 e.g.~\cite{Yang:2016zbm,Yang:2017bls}, it
 is convenient to define the extragalactic (or excess) DM of FRB as
 \be{eq6}
 \rm DM_E\equiv DM_{obs}-DM_{MW}=DM_{IGM}+DM_{HG}\,.
 \ee

Actually, the main contribution to DM of FRB comes from
 IGM, which can be obtained by using Eq.~(\ref{eq4}). In
 e.g.~\cite{Ioka:2003fr,Inoue:2003ga}, $\rm DM_{IGM}$ for a
 fully ionized and pure hydrogen plasma has been studied.
 Assuming that all baryons are fully ionized and homogeneously
 distributed, the number density of free electrons is given
 by~\cite{Ioka:2003fr,Inoue:2003ga}
 \be{eq7}
 n_{e,\,z}=n_{b,\,z}=\rho_{b,\,z}/m_p=\rho_{b,\,0}\,(1+z)^3/m_p
 =\frac{3H_0^2\,\Omega_{b,\,0\,}(1+z)^3}{8\pi Gm_p}\,,
 \ee
 where $\Omega_{b,\,0}= 8\pi G\rho_{b,\,0}/(3H_0^2)$ is the
 well-known present fractional density of baryons
 (the subscript ``\,0\,'' indicates the present value of the
 corresponding quantity), $H_0$ is the Hubble constant,
 $m_p$ is the mass of proton. On the other hand, the element
 of distance~\cite{Ioka:2003fr,Inoue:2003ga}
 \be{eq8}
 dl=c\,dt=c\,dz\left|\frac{dt}{dz}\right|=\frac{c\,dz}{(1+z)\,H(z)}\,,
 \ee
 where $H\equiv\dot{a}/a$ is the Hubble parameter, $a=(1+z)^{-1}$ is
 the scale factor, and a dot denotes the derivative with respect to
 cosmic time $t$. Using Eqs.~(\ref{eq7}), (\ref{eq8}) and
 (\ref{eq4}), one obtain~\cite{Ioka:2003fr,Inoue:2003ga}
 \be{eq9}
 \langle{\rm DM_{IGM}}\rangle=\frac{3cH_0\Omega_{b,\,0}}{8\pi G m_p}
 \int_0^z\frac{(1+\tilde{z})\,d\tilde{z}}{E(\tilde{z})}\,,
 \ee
 where $E\equiv H/H_0$, and $\langle{\rm DM_{IGM}}\rangle$ is
 the mean of $\rm DM_{IGM}$, while $\rm DM_{IGM}$ will deviate
 from $\langle{\rm DM_{IGM}}\rangle$ if the plasma density
 fluctuations are taken into account~\cite{McQuinn:2013tmc}
 (see also~\cite{Ioka:2003fr,Jaroszynski:2018vgh}). However,
 this is the simplified case. A more general and realistic case was
 discussed in~\cite{Deng:2013aga}, by considering IGM comprised
 of hydrogen and helium which might be not fully ionized. The
 hydrogen (H) mass fraction $Y_{\rm H}=(3/4)\,y_1$, and the
 helium (He) mass fraction $Y_{\rm He}=(1/4)\,y_2$, where
 $y_1\sim 1$ and $y_2\simeq 4-3y_1\sim 1$ are the hydrogen and
 helium mass fractions normalized to the typical values $3/4$
 and $1/4$, respectively. Their ionization
 fractions $\chi_{e,\,\rm H}(z)$ and $\chi_{e,\,\rm He}(z)$ are
 functions of redshift $z$. Noting that H and He have 1 and 2
 electrons respectively, the number density of free electrons
 at redshift $z$ is given by~\cite{Deng:2013aga,Yang:2016zbm}
 \bea
 n_{e,\,z}&=&n_{{\rm H},\,z}\,\chi_{e,\,\rm H}(z)
 +2\,n_{{\rm He},\,z}\,\chi_{e,\,\rm He}(z)=
 \left[\,n_{{\rm H},\,0}\,\chi_{e,\,\rm H}(z)+
 2\,n_{{\rm He},\,0}\,\chi_{e,\,\rm He}(z)\,\right](1+z)^3\nonumber\\[2mm]
 &=&\left[\,\frac{Y_{\rm H}\,\rho_{b,\,0}f_{\rm IGM}}{m_p}\,
 \chi_{e,\,\rm H}(z)+2\,\frac{Y_{\rm He}\,\rho_{b,\,0}
 f_{\rm IGM}}{4m_p}\,\chi_{e,\,\rm He}(z)\,\right](1+z)^3\nonumber\\[1.5mm]
 &=&\frac{3H_0^2\,\Omega_{b,\,0}f_{\rm IGM}}{8\pi G m_p}\,f_e(z)\cdot
 (1+z)^3\,,\label{eq10}
 \eea
 where $f_{\rm IGM}$ is the fraction of baryon mass in
 the intergalactic medium, and
 \be{eq11}
 f_e(z)\equiv\frac{3}{4}\,y_1\,\chi_{e,\,\rm H}(z)+\frac{1}{8}
 \,y_2\,\chi_{e,\,\rm He}(z)\,.
 \ee
 Using Eqs.~(\ref{eq10}), (\ref{eq8}) and (\ref{eq4}),
 one obtain~\cite{Deng:2013aga,Yang:2016zbm}
 \be{eq12}
 \langle{\rm DM_{IGM}}\rangle=K_{\rm IGM}\int_0^z\frac{f_e(\tilde{z})
 \cdot(1+\tilde{z})\,d\tilde{z}}{E(\tilde{z})}\,,
 \ee
 where
 \be{eq13}
 K_{\rm IGM}\equiv\frac{3cH_0\Omega_{b,\,0}f_{\rm IGM}}{8\pi G m_p}\,.
 \ee
 Obviously, $\langle{\rm DM_{IGM}}\rangle$ in Eq.~(\ref{eq12})
 is the generalized version of the one in Eq.~(\ref{eq9}).
 The current observations suggest that the intergalactic
 hydrogen and helium are fully ionized at redshift $z\lesssim 6$ and
 $z\lesssim 3$~\cite{Meiksin:2007rz,Becker:2010cu}, respectively. Therefore,
 following e.g.~\cite{Yang:2016zbm,Yang:2017bls}, in the present work
 we only consider FRBs at redshift $z\leq 3$ to ensure that
 hydrogen and helium are both fully ionized, and hence
 $\chi_{e,\,\rm H}(z)=\chi_{e,\,\rm He}(z)=1$. In this case,
 $f_e(z)\simeq 7/8$. Following e.g.~\cite{Yang:2016zbm,Yang:2017bls,
 Gao:2014iva}, we adopt $f_{\rm IGM}=0.83$ (see
 e.g.~\cite{Fukugita:1997bi,Shull:2011aa} and \cite{Deng:2013aga}).

In this work, we employ Monte Carlo simulations of FRBs to test
 the possible cosmic anisotropy. Here, we generate the simulated FRBs
 by using the simplest flat $\Lambda$CDM model as the fiducial
 cosmology. As is well known, in this case, the dimensionless Hubble
 parameter is given by
 \be{eq14}
 E(z)=\left[\,\Omega_{m,\,0}(1+z)^3+(1-\Omega_{m,\,0})\,\right]^{1/2}\,,
 \ee
 where $\Omega_{m,\,0}$ is the present fractional density of
 matter. We adopt the latest flat $\Lambda$CDM parameters
 from Planck 2018 CMB data~\cite{Aghanim:2018eyx}, namely
 $H_0=67.36\;{\rm km/s/Mpc}$, $\Omega_{m,\,0}=0.3153$, and
 $\Omega_{b,\,0}=0.0493$. In this case, $K_{\rm IGM}=928.0118\;
 {\rm pc\hspace{0.24em} cm^{-3}}$. With these fiducial parameters, we can
 get the mean $\langle{\rm DM_{IGM}}\rangle$ by using Eq.~(\ref{eq12}).
 As mentioned above, $\rm DM_{IGM}$ deviates from
 $\langle{\rm DM_{IGM}}\rangle$ if the plasma density fluctuations are
 taken into account~\cite{McQuinn:2013tmc} (see
 also e.g.~\cite{Ioka:2003fr,Jaroszynski:2018vgh}). According
 to~\cite{McQuinn:2013tmc}, $\rm DM_{IGM}$ can be approximated
 by a Gaussian distribution with a random fluctuation $\sigma_{\rm IGM}
 =100\;{\rm pc\hspace{0.24em} cm^{-3}}$~\cite{McQuinn:2013tmc} (see also
 e.g.~\cite{Yang:2016zbm,Ioka:2003fr,Jaroszynski:2018vgh}),
 \be{eq15}
 {\rm DM_{IGM}}={\cal N}\left(\langle{\rm DM_{IGM}}\rangle,
 \,\sigma_{\rm IGM}\right)\,.
 \ee

On the other hand, $\rm DM_{HG}$, namely the contribution from
 host galaxy of FRB to DM, is poorly known. It depends on many
 factors, such as the type of host galaxy, the site of FRB in
 host galaxy, and the inclination angle of the disk with respect to
 line of sight~\cite{Gao:2014iva}. The local DMs of FRB host
 galaxies might be assumed to have no significant evolution
 with redshift~\cite{Yang:2017bls}, namely the
 mean $\langle{\rm DM_{HG,\,loc}}\rangle\sim{\rm const.}$, and
 $\rm DM_{HG,\,loc}$ can be approximated by a Gaussian distribution
 with a random fluctuation $\sigma_{\rm HG,\,loc}$~\cite{Yang:2016zbm,
 Gao:2014iva,Yang:2017bls,Zhou:2014yta}, namely
 \be{eq16}
 {\rm DM_{HG,\,loc}}={\cal N}\left(\langle{\rm DM_{HG,\,loc}}\rangle,
 \,\sigma_{\rm HG,\,loc}\right)\,.
 \ee
 To determine the values of $\langle{\rm DM_{HG,\,loc}}\rangle$
 and $\sigma_{\rm HG,\,loc}\,$, it is helpful to consult our
 Galaxy (Milky Way). As mentioned above, $\disp \rm DM_{MW}
 \lesssim 100\;{\rm pc\hspace{0.24em} cm^{-3}}$ at high Galactic latitude
 $|b|>10^\circ$, and its average dispersion is a few tens of
 $\rm pc\hspace{0.24em} cm^{-3}$~\cite{Taylor:1993my,Manchester:2004bp}
 (see also e.g.~\cite{Gao:2014iva,Zhou:2014yta}). So, it
 is reasonable to adopt the fiducial values $\langle{\rm DM_{HG,\,loc}}
 \rangle=100\;{\rm pc\hspace{0.24em} cm^{-3}}$ and $\sigma_{\rm HG,\,loc}=
 20\;{\rm pc\hspace{0.24em} cm^{-3}}$ following e.g.~\cite{Yang:2016zbm}.
 For a FRB at redshift~$z$, its observed $\rm DM_{HG}$ and the uncertainty
 should be redshifted (see e.g.~\cite{Yang:2016zbm,Gao:2014iva,Yang:2017bls,
 Zhou:2014yta}),
 \be{eq17}
 {\rm DM_{HG}}={\rm DM_{HG,\,loc}}/(1+z)\,,\quad\quad\quad
 \sigma_{\rm HG}=\sigma_{\rm HG,\,loc}/(1+z)\,.
 \ee

There is no existing guideline for the redshift distribution
 of FRBs by now. Following e.g.~\cite{Yang:2016zbm,Gao:2014iva,
 Zhou:2014yta}, we assume that the redshift distribution
 of FRBs takes a form similar to the one of GRBs~\cite{Shao:2011xt},
 \be{eq18}
 P(z)\propto z e^{-z}\,.
 \ee
 For a simulated FRB, we can randomly assign a redshift $z$
 from this distribution. Then, the corresponding
 $\langle{\rm DM_{IGM}}\rangle$ can be obtained by using
 Eq.~(\ref{eq12}), and hence we can assign a random value to
 $\rm DM_{IGM}$ from the Gaussian distribution (\ref{eq15}).
 On the other hand, the value of $\rm DM_{HG,\,loc}$ can be
 assigned randomly from the Gaussian distribution (\ref{eq16}).
 Finally, the extragalactic (or excess) DM
 defined in Eq.~(\ref{eq6}) of this simulated FRB is given by
 \be{eq19}
 {\rm DM_E=DM_{IGM}+DM_{HG,\,loc}}/(1+z)\,.
 \ee


 \begin{center}
 \begin{figure}[tb]
 \centering
 \vspace{-4mm}  
 \includegraphics[width=1.0\textwidth]{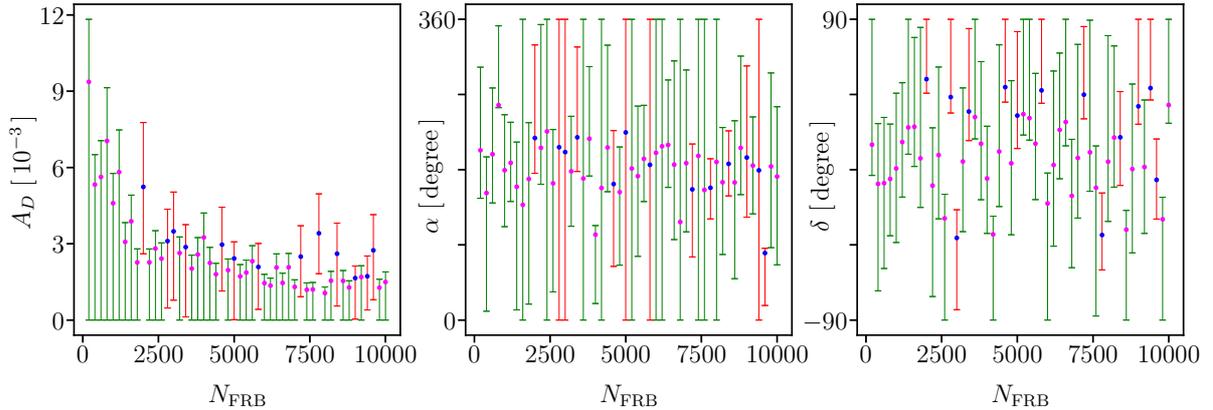}
 \caption{\label{fig1} The marginalized $1\sigma$ constraints
 on the amplitude $A_D$, the right ascension $\alpha$, and the
 declination $\delta$ of the dipole, by using various simulated
 datasets consisting of $N_{\rm FRB}$ FRBs generated with a
 preset dipole $A_D^{\rm fid}=10^{-3}$, $\alpha_{\rm fid}=150^\circ$,
 and $\delta_{\rm fid}=10^\circ$. The green error bars with magenta
 means (the red error bars with blue means) indicate the cases that
 $A_D=0$ is consistent (inconsistent) with the simulated FRB dataset
 in the $1\sigma$ region, respectively. Note that $A_D$ is given in
 units of $10^{-3}$. See the text for details.}
 \end{figure}
 \end{center}


\vspace{-11mm}  


\section{Testing the cosmic anisotropy with the simulated FRBs}\label{sec3}


\subsection{Generating the simulated FRB datasets with a
 preset direction}\label{sec3a}

In fact, the methodology to simulate FRBs given in Sec.~\ref{sec2} is
 statistically isotropic. Although there exists variance in $\rm DM_E$
 for different lines of sight, it is actually statistical noise due
 to random fluctuations, and hence there is no preferred direction in
 the simulated FRB datasets indeed.

There exist various methods to generate the simulated datasets
 with a preset direction in the literature (e.g.~\cite{Deng:2018yhb,
 Lin:2016jqp,Cai:2017aea,Lin:2018azu}). A simple way is to directly
 put a dipole with the preset direction into the simulated data under
 consideration. In our case, the simulated data of FRB is the
 extragalactic (or excess) dispersion measure $\rm DM_E$, similar to
 the cases considered in e.g.~\cite{Yang:2016zbm}. Since the
 simulated $\rm DM_E$ given in Eq.~(\ref{eq19})
 is statistically isotropic in fact, we refer to it as
 $\rm DM_E^{iso}$ instead. The simulated $\rm DM_E$ with a
 preset direction $\hat{n}_{\rm fid}$ is given by
 \be{eq20}
 {\rm DM_E}={\rm DM_E^{iso}}\left[\,1+A_D^{\rm fid}\left(
 \hat{n}_{\rm fid}\cdot\hat{p}\right)\,\right]\,,
 \ee
 where $\rm DM_E^{iso}$ is actually the one given
 in Eq.~(\ref{eq19}), $A_D^{\rm fid}$ is the amplitude of the
 preset fiducial dipole. The preset fiducial dipole direction
 $\hat{n}_{\rm fid}$ in terms of the equatorial coordinates
 $(\alpha,\,\delta)$ is given by
 \be{eq21}
 \hat{n}_{\rm fid}=\cos(\delta_{\rm fid})\cos(\alpha_{\rm fid})
 \;\hat{\bf i} + \cos(\delta_{\rm fid})\sin(\alpha_{\rm fid})
 \;\hat{\bf j} + \sin(\delta_{\rm fid})\;\hat{\bf k}\,,
 \ee
 where $\hat{\bf i}$, $\hat{\bf j}$, $\hat{\bf k}$ are the unit
 vectors along the axes of Cartesian coordinate system, and
 $\alpha_{\rm fid}$, $\delta_{\rm fid}$ are the right ascension (ra),
 declination (dec) of the preset fiducial dipole direction,
 respectively. The position of the $i$-th simulated data point with
 the equatorial coordinates $(\alpha_i,\,\delta_i)$ is given by
 \be{eq22}
 \hat{p}_i=\cos(\delta_i)\cos(\alpha_i)\;\hat{\bf i}+\cos(\delta_i)
 \sin(\alpha_i)\;\hat{\bf j}+\sin(\delta_i)\;\hat{\bf k}\,.
 \ee
 Note that in the present work, we arbitrarily adopt the preset
 fiducial dipole direction as $\alpha_{\rm fid}=150^\circ$ and
 $\delta_{\rm fid}=10^\circ$. On the other hand, the amplitude
 of the preset fiducial dipole $A_D^{\rm fid}$
 will be specified in the particular simulation (see below).

Let us briefly describe the main steps to generate the
 simulated FRB datasets with a preset direction:

\begin{enumerate}[(A)]
  \vspace{-0.6\itemsep}  
  \setlength{\itemindent}{1.8em}
  \setlength{\itemsep}{0.4\itemsep}

  \item Assign a random number uniformly taken
  from $[\,0^\circ,\,360^\circ)$ to the simulated FRB as its
  right ascension $\alpha_i\,$, and assign a random number
  uniformly taken from $[\,-90^\circ,\,+90^\circ\,]$ to this
  simulated FRB as its declination $\delta_i\,$.

  \item Assign a random redshift $z_i\leq 3$ from
  the distribution in Eq.~(\ref{eq18}) to this simulated FRB.

  \item Calculate ${\rm DM_{E,\,{\it i}}^{iso}=
  DM_{IGM,\,{\it i}}+DM_{HG,\,loc,\,{\it i}}}/(1+z_i)$ as
  described in Sec.~\ref{sec2}.

  \item Generate $\rm DM_{E,\,{\it i}}$ for this simulated FRB
  by using Eqs.~(\ref{eq20})$\,\sim\,$(\ref{eq22}) with
  a preset dipole direction.

  \item Generate the error $\sigma_{{\rm E},\,i}=\big\{
  \sigma_{\rm IGM}^2+\left[\,\sigma_{\rm HG,\,loc}/(1+z_i)\,
  \right]^2\big\}^{1/2}$ for this FRB data point.

  \item Repeat the above steps for $N_{\rm FRB}$ times
  to generate $N_{\rm FRB}$ simulated FRBs.

\end{enumerate}

\noindent The formatted output data file for the simulated FRBs
 contains $N_{\rm FRB}$ rows of $\left\{\alpha_i,\,\delta_i,\,z_i,\,
 {\rm DM_{E,\,{\it i}}},\,\sigma_{{\rm E},\,i}\right\}$. Once a simulated
 FRB dataset has been generated, one should forget everything used to
 generate it, including all the fiducial cosmology, parameters, and preset
 dipole. One should pretend to deal with it as a real ``\,observational\,''
 dataset blindly.


 \begin{center}
 \begin{figure}[tb]
 \centering
 \vspace{-1.5mm}  
 \includegraphics[width=1.0\textwidth]{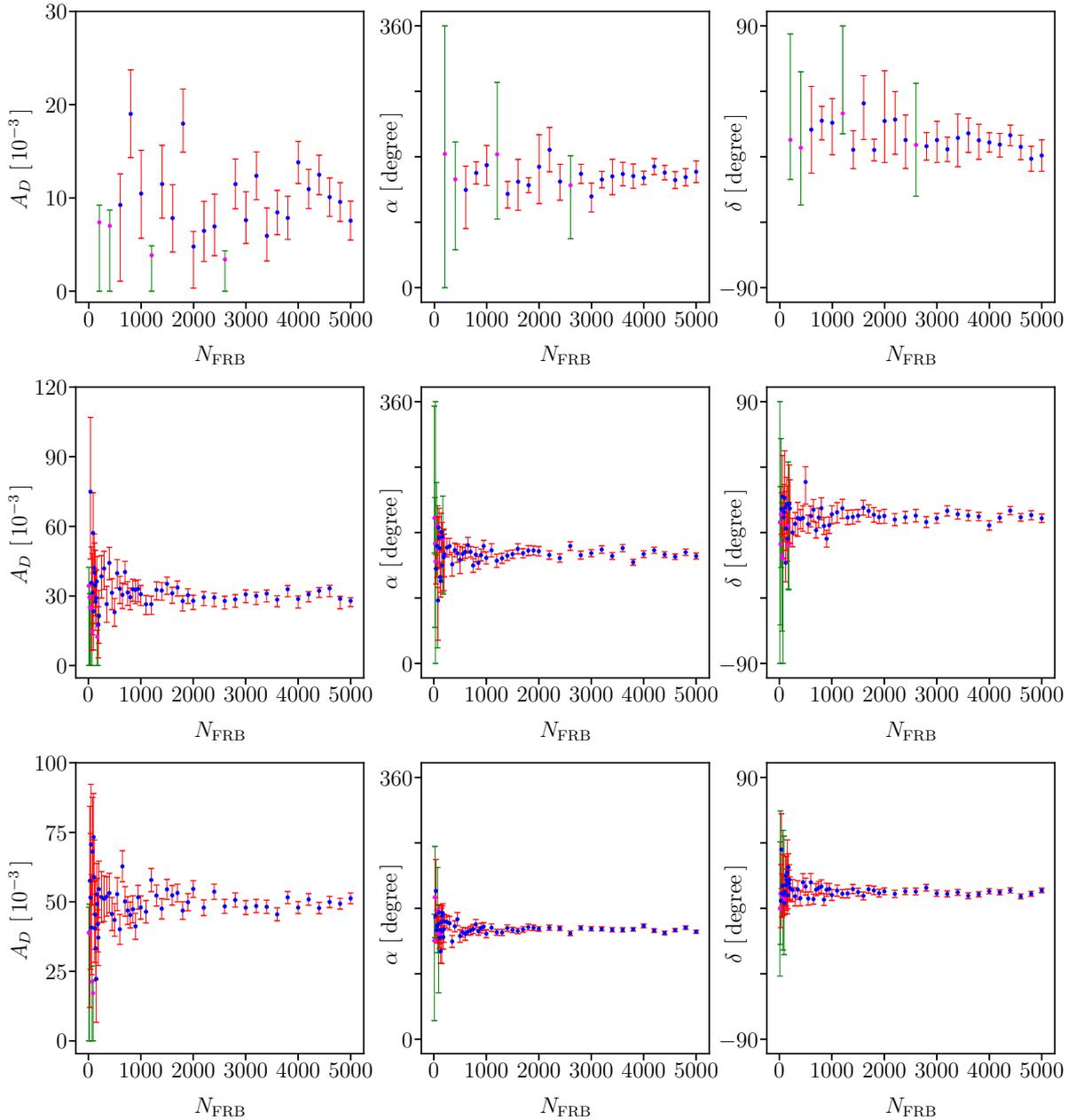}
 \caption{\label{fig2} The same as in Fig.~\ref{fig1}, except
 for $A_D^{\rm fid}=0.01$ (top panels), $0.03$ (middle panels)
 and $0.05$ (bottom panels), respectively. See the text for
 details.}
 \end{figure}
 \end{center}



 \begin{center}
 \begin{figure}[tb]
 \centering
 \vspace{-9mm}  
 \includegraphics[width=0.8\textwidth]{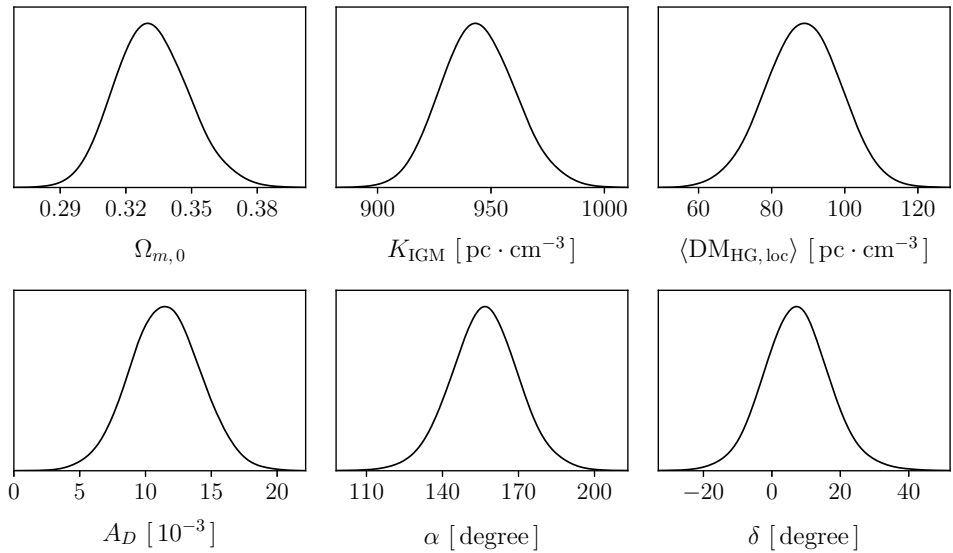}
 \caption{\label{fig3} The marginalized probability distributions of
 all the 6 free model parameters
 $\Omega_{m,\,0}$, $K_{\rm IGM}$, $\langle{\rm DM_{HG,\,loc}}\rangle$,
 $A_D$, $\alpha$, $\delta$, by using 2800 FRBs simulated with
 $A_D^{\rm fid}=0.01$. Note that $A_D$ is given in units
 of $10^{-3}$. See the text and the top panels of Fig.~\ref{fig2} for
 details.}
 \end{figure}
 \end{center}



 \begin{center}
 \begin{figure}[tb]
 \centering
 \vspace{-4mm}  
 \includegraphics[width=1.0\textwidth]{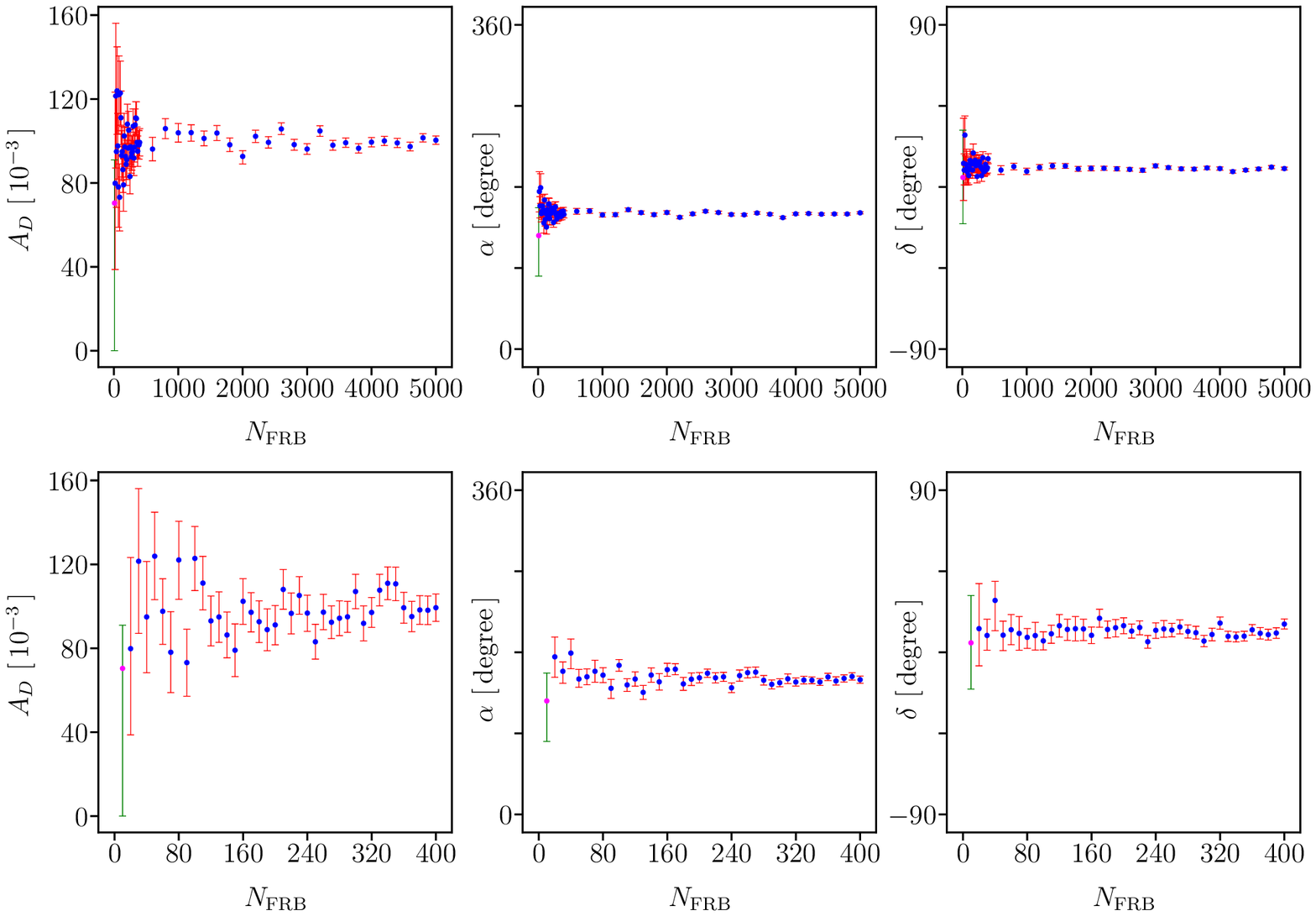}
 \caption{\label{fig4} The same as in Fig.~\ref{fig1}, except
 for $A_D^{\rm fid}=0.1$. The bottom panels are actually the
 foreparts $N_{\rm FRB}\leq 400$ of the top panels. See the
 text for details.}
 \end{figure}
 \end{center}


\vspace{-29mm}  


\subsection{Testing the cosmic anisotropy with
 the simulated FRB datasets}\label{sec3b}

Here, we try to test the possible cosmic anisotropy with the
 simulated FRB datasets. We assume that the universe can be
 theoretically described by a flat $\Lambda$CDM model, and
 the corresponding dimensionless Hubble parameter $E(z)$ is
 given in Eq.~(\ref{eq14}). We consider the extragalactic (or
 excess) dispersion measure $\rm DM_E$ with a possible dipole,
 \be{eq23}
 {\rm DM_E^{th}}=\langle{\rm DM_E}\rangle\left[\,1+A_D
 \left(\hat{n}\cdot\hat{p}\right)\,\right]\,,
 \ee
 where $\langle{\rm DM_E}\rangle=\langle{\rm DM_{IGM}}\rangle
 +\langle{\rm DM_{HG,\,loc}}\rangle/(1+z)$, and
 $\langle{\rm DM_{IGM}}\rangle$ is given in Eq.~(\ref{eq12}).
 The dipole direction $\hat{n}$ in terms of the equatorial
 coordinates $(\alpha,\,\delta)$ is given by
 \be{eq24}
 \hat{n}=\cos(\delta)\cos(\alpha)\;\hat{\bf i}+
 \cos(\delta)\sin(\alpha)\;\hat{\bf j}+\sin(\delta)\;\hat{\bf k}\,.
 \ee
 There are 6 free model parameters, namely $\Omega_{m,\,0}$,
 $K_{\rm IGM}$, $\langle{\rm DM_{HG,\,loc}}\rangle$, $A_D$,
 $\alpha$ and $\delta$. The constraints on these 6 free model
 parameters can be obtained by using the simulated FRB
 dataset. The corresponding $\chi^2$ is given by
 \be{eq25}
 \chi^2\left(\Omega_{m,\,0},\,K_{\rm IGM},\,
 \langle{\rm DM_{HG,\,loc}}\rangle,\,A_D,\,\alpha,\,\delta
 \right)=\sum_i\frac{\left({\rm DM}_{{\rm E},\,i}
 -{\rm DM_E^{th}}\right)^2}{\sigma_{{\rm E},\,i}^2}\,.
 \ee
 In the following, we use the Markov Chain Monte Carlo (MCMC)
 code CosmoMC~\cite{Lewis:2002ah} to this end. Noting that
 $A_D\left(\hat{n}\cdot\hat{p}\right)=-A_D\left(-\hat{n}\cdot
 \hat{p}\right)$ in Eq.~(\ref{eq23}), a positive $A_D$ with a
 direction $\hat{n}$ is equivalent to a negative $A_D$ with an
 opposite direction $-\hat{n}$. Therefore, in this work,
 without loss of generality, we require $A_D\geq 0$ as prior
 when running CosmoMC.

It is of interest to see how many FRBs (at least) are required
 to find a possible cosmic anisotropy. At first, we consider
 a tiny cosmic anisotropy represented by a dipole with
 $A_D^{\rm fid}=0.001$. Adopting this $A_D^{\rm fid}$, we
 generate a series of simulated FRB datasets consisting of
 $N_{\rm FRB}=200$, 400, 600, ..., 10000 FRBs, respectively.
 For each simulated FRB dataset, we can obtain the constraints
 on the 6 free model parameters mentioned above, by using the
 MCMC code CosmoMC. We focus on the parameters related to the
 dipole, namely $A_D$, $\alpha$ and $\delta$. We present
 the marginalized $1\sigma$ constraints on these 3 dipole parameters
 versus $N_{\rm FRB}$ in Fig.~\ref{fig1}. The green error bars with
 magenta means (the red error bars with blue means) indicate
 the cases that $A_D=0$ is consistent (inconsistent) with the
 simulated FRB dataset in the $1\sigma$ region, respectively.
 If $A_D=0$ is consistent with the ``\,observational\,'' dataset, it
 means that no preferred direction is found. Unfortunately,
 from Fig.~\ref{fig1}, we see that even up to the case of
 $N_{\rm FRB}=10000$, $A_D=0$ is still consistent with the
 simulated FRB dataset. On the other hand, the constraints on
 the dipole direction $(\alpha,\,\delta)$ are fairly loose.
 Even in the cases that $A_D=0$ is not included in the
 $1\sigma$ region, the ``\,found\,'' $1\sigma$ angular regions
 are too wide to say that a preferred direction is really
 found. Thus, FRBs are not competent to find the tiny cosmic
 anisotropy with a dipole amplitude of ${\cal O}(10^{-3})$.
 We will come back to this issue in the next subsection
 (Sec.~\ref{sec3c}).

We turn to the case of a larger cosmic anisotropy represented
 by a dipole with $A_D^{\rm fid}=0.01$. Adopting
 this $A_D^{\rm fid}$, we generate a series of simulated FRB
 datasets consisting of $N_{\rm FRB}=200$, 400, 600, ..., 5000 FRBs,
 respectively. Similarly, we present the marginalized $1\sigma$
 constraints on the dipole parameters $A_D$, $\alpha$, $\delta$
 versus $N_{\rm FRB}$ in the top panels of Fig.~\ref{fig2}. It
 is easy to see that in all cases of $N_{\rm FRB}\geq 2800$, a
 non-zero $A_D$ beyond $1\sigma$ region can be found, while the
 $1\sigma$ constraints on $\alpha$ and $\delta$ are also tight.
 Therefore, at least 2800 FRBs are competent to find the cosmic
 anisotropy with $A_D^{\rm fid}=0.01$. More FRBs lead to
 tighter constraints on the cosmic anisotropy. It is of
 interest to see also the constraints on the other free model
 parameters. In Fig.~\ref{fig3}, we present the marginalized
 probability distributions of all the 6 free model parameters
 for the case of $N_{\rm FRB}=2800$. Obviously, the constraints
 on all the 6 parameters are consistent with the fiducial
 ones used to generate this simulated FRB dataset. For
 conciseness, we choose not to show the constraints on all the
 6 parameters again in the rest of this paper, since we
 are mainly interested in the 3 parameters related to the
 cosmic anisotropy, namely $A_D$, $\alpha$ and $\delta$.

For the case of $A_D^{\rm fid}=0.03$, we also present the
 marginalized $1\sigma$ constraints on the dipole parameters
 $A_D$, $\alpha$, $\delta$ versus $N_{\rm FRB}=10$, 20, 30,
 ..., 180, 190, 200, 250, 300, 350, 400, ..., 950, 1000, 1100,
 ..., 2000, 2200, ..., 5000 in the middle panels of Fig.~\ref{fig2}.
 We see that in all cases of $N_{\rm FRB}\geq 190$, a non-zero $A_D$
 beyond $1\sigma$ region can be found, while the $1\sigma$ constraints
 on $\alpha$ and $\delta$ are fairly tight. Obviously, for all cases
 of $N_{\rm FRB}\geq 190$, the constraints on the parameters $A_D$,
 $\alpha$, $\delta$ are well consistent with the fiducial ones used
 to generate the simulated FRB datasets. Therefore, at least
 190 FRBs are competent to find the cosmic anisotropy with
 $A_D^{\rm fid}=0.03$.

Similarly, we present the corresponding results for the case of
 $A_D^{\rm fid}=0.05$ in the bottom panels of Fig.~\ref{fig2}.
 Obviously, the constraints on the parameters $A_D$, $\alpha$,
 $\delta$ become tighter (especially for the angular parameters
 $\alpha$ and $\delta$). In all cases of $N_{\rm FRB}\geq 100$,
 a non-zero $A_D$ beyond $1\sigma$ region can be found, while
 the $1\sigma$ constraints on $\alpha$ and $\delta$ are fairly
 tight. In other words, at least 100 FRBs are competent to find
 the cosmic anisotropy with $A_D^{\rm fid}=0.05$.

Finally, we consider a large cosmic anisotropy with $A_D^{\rm fid}=0.1$,
 and present the corresponding results in Fig.~\ref{fig4}. To see clearly,
 we also enlarge the parts of $N_{\rm FRB}\leq 400$ in the bottom panels
 of Fig.~\ref{fig4}. For such a big cosmic anisotropy, it is easy to
 find the non-zero dipole with high precision by using very few FRBs
 with known redshifts. In fact, at least 20 FRBs with known redshifts
 are competent to find the cosmic anisotropy with $A_D^{\rm fid}=0.1$.
 However, by now, there are only a few FRBs (e.g. the repeater FRB 121102)
 with identified redshifts, and hence the published FRBs to date are still
 not enough. We expect that such a big cosmic anisotropy can be ruled out
 by using only a few tens of FRBs with known redshifts in the near future.


 \begin{center}
 \begin{figure}[tb]
 \centering
 \vspace{-3mm}  
 \includegraphics[width=1.0\textwidth]{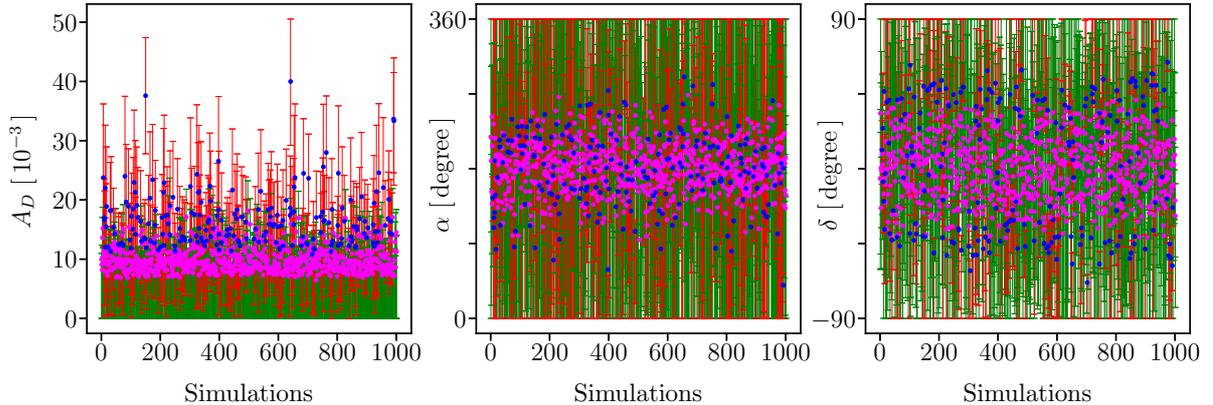}
 \caption{\label{fig5} The marginalized $1\sigma$ constraints
 on the amplitude $A_D$, the right ascension $\alpha$, and the
 declination $\delta$ of the dipole, by using 1000 simulated
 datasets consisting of $N_{\rm FRB}=200$ FRBs generated
 without a preset dipole (namely $A_D^{\rm fid}=0$). The green
 error bars with magenta means (the red error bars with blue
 means) indicate the cases that $A_D=0$ is consistent (inconsistent)
 with the simulated FRB dataset in the $1\sigma$ region,
 respectively. Note that $A_D$ is given in units
 of $10^{-3}$. See the text for details.}
 \end{figure}
 \end{center}


\vspace{-12.5mm}  


\subsection{Ratio of pseudo anisotropic signals from the
 statistical noise}\label{sec3c}

An important question is how reliable are the above results?
 In fact, pseudo anisotropic signals from the statistical noise
 due to random fluctuations are possible. Here, we would like
 to test this possibility in more details.

The key is to find the pseudo anisotropic signal in the
 simulated FRB datasets generated without a preset anisotropy
 (namely $A_D^{\rm fid}=0$). Noting that at least 190 FRBs are
 competent to find a cosmic dipole with amplitude $\sim 0.03$
 as mentioned above, we randomly generate 1000 simulated
 datasets consisting of $N_{\rm FRB}=200$ FRBs without a preset
 dipole (namely $A_D^{\rm fid}=0$). For each simulated dataset
 consisting of $N_{\rm FRB}=200$ FRBs, we can obtain the
 constraints on the 6 free model parameters, following the
 same procedures used in the previous subsection, as if
 $A_D^{\rm fid}\not=0$. In Fig.~\ref{fig5}, we present the
 marginalized $1\sigma$ constraints on the 3 dipole parameters
 for each simulated dataset consisting of $N_{\rm FRB}=200$
 FRBs. In all the 1000 simulations, there are 212 simulations
 having a non-zero $A_D$ beyond $1\sigma$ region, as shown by
 the red error bars with blue means in the left panel of
 Fig.~\ref{fig5}. However, a non-zero $A_D$ is not enough to
 say that a preferred direction has been found. In fact, many
 of them correspond to a very wide $1\sigma$ angular region,
 namely the $1\sigma$ constraints on the angular parameters
 $\alpha$ and $\delta$ are very loose, as shown by the long
 red error bars with blue means in the middle and right panels
 of Fig.~\ref{fig5}. In some cases, the corresponding direction
 can be the whole sky or a half sky. Therefore, we cannot say
 that a preferred direction has been really found. On the
 contrary, we would like to mention the fairly tight $1\sigma$
 constraints on $\alpha$ and $\delta$ in the simulations
 generated with a no-zero dipole ($A_D^{\rm fid}\not=0$), as
 shown in the the middle and right panels of Figs.~\ref{fig2}
 and \ref{fig4}. Here, we propose a fairly loose direction
 criterion, namely the $1\sigma$ range (upper bound minus lower
 bound) of the right ascension $\alpha$ should be $\leq 180^\circ$,
 and the $1\sigma$ range (upper bound minus lower bound)
 of the declination $\delta$ should be $\leq 90^\circ$.
 Noting that the value ranges of $\alpha$ and $\delta$ are
 $[\,0^\circ,\,360^\circ)$ and $[\,-90^\circ,\,+90^\circ\,]$
 respectively, this direction criterion is indeed fairly loose.
 Actually, in the 212 simulations with a non-zero $A_D$ beyond
 $1\sigma$ region mentioned above, only 115 simulations can
 pass this loose direction criterion. However, only 30 of these
 115 simulations have a $1\sigma$ upper bound on $A_D$ higher
 than $0.03$ (please remember that 200 FRBs are competent to
 find a cosmic dipole with amplitude $\sim 0.03$ as mentioned
 above). In other words, when we report a cosmic dipole with
 amplitude $\sim 0.03$ by using 200 FRBs, there are only 30
 pseudo anisotropic signals from the statistical noise in 1000
 simulations. The ratio of pseudo anisotropic signals from the
 statistical noise is around $30/1000=3\%$, which is acceptable
 in fact. Note that the ratio of pseudo anisotropic signals
 will significantly decrease with more FRBs (for example, when
 one reports a cosmic dipole with amplitude $\sim 0.03$ by
 using $500+$ FRBs with known redshifts).

\begin{center}
 \begin{figure}[tb]
 \centering
 \vspace{-3mm}  
 \includegraphics[width=1.0\textwidth]{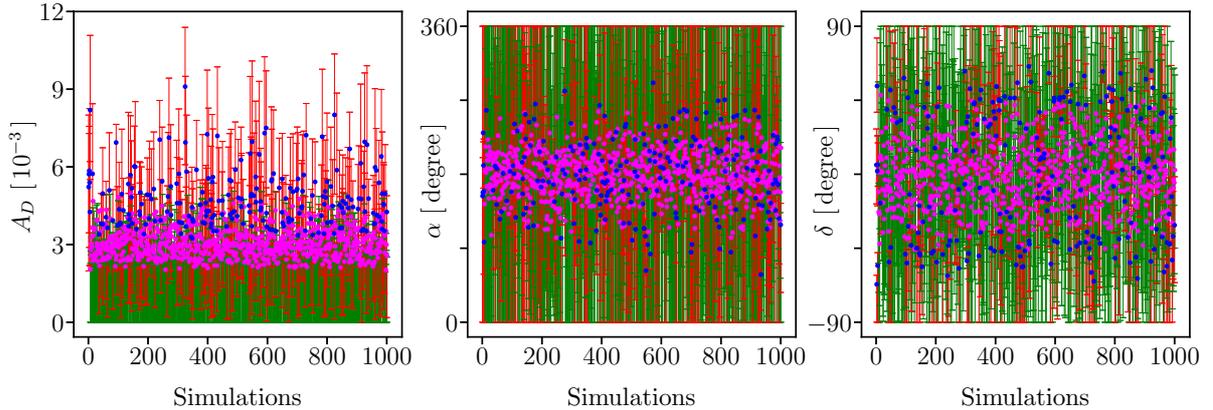}
 \caption{\label{fig6} The same as in Fig.~\ref{fig5}, except
 for 1000 simulated datasets consisting of $N_{\rm FRB}=3000$
 FRBs generated without a preset dipole (namely
 $A_D^{\rm fid}=0$). See the text for details.}
 \end{figure}
 \end{center}

Similarly, noting that at least 2800 FRBs are competent to find
 a cosmic dipole with amplitude $\sim 0.01$ as mentioned above,
 we randomly generate 1000 simulated datasets consisting of
 $N_{\rm FRB}=3000$ FRBs without a preset dipole (namely
 $A_D^{\rm fid}=0$). In Fig.~\ref{fig6}, we present the marginalized
 $1\sigma$ constraints on the 3 dipole parameters for each simulated
 dataset consisting of $N_{\rm FRB}=3000$ FRBs. In all the 1000
 simulations, there are 218 simulations having a non-zero $A_D$
 beyond $1\sigma$ region, as shown by the red error bars with
 blue means in the left panel of Fig.~\ref{fig6}. Again, many
 of them correspond to a very wide $1\sigma$ angular region,
 namely the $1\sigma$ constraints on the angular parameters
 $\alpha$ and $\delta$ are very loose, as shown by the long
 red error bars with blue means in the middle and right panels
 of Fig.~\ref{fig6}. In these 218 simulations with a non-zero
 $A_D$ beyond $1\sigma$ region mentioned above, only 133 simulations
 can pass the loose direction criterion proposed above. However, only
 3 of these 133 simulations have a $1\sigma$ upper bound on
 $A_D$ higher than $0.01$ (please remember that 3000 FRBs are
 competent to find a cosmic dipole with amplitude $\sim 0.01$
 as mentioned above). Thus, when we report a cosmic dipole with
 amplitude $\sim 0.01$ by using 3000 FRBs, the ratio of pseudo
 anisotropic signals from the statistical noise is around
 $3/1000=0.3\%$.

Through the above two concrete examples, we show that the
 results obtained in Sec.~\ref{sec3b} are reliable, because the
 ratio of pseudo anisotropic signals from the statistical noise
 is fairly low. It is worth noting that the available FRBs with
 known redshift will be numerous in the coming years, as
 mentioned in Sec.~\ref{sec1}. With numerous FRBs, it is
 reasonable to expect that the ratio of pseudo anisotropic
 signals from the statistical noise will become much less than
 $0.1\%$.

As a byproduct, from the above simulations, we can also
 understand why FRBs are not competent to find the tiny cosmic
 anisotropy with a dipole amplitude of ${\cal O}(10^{-3})$,
 even by using 10000 FRBs with known redshifts, as mentioned
 in Sec.~\ref{sec3b}. As shown in the left panels
 of Figs.~\ref{fig5} and \ref{fig6}, most of the means (magenta
 and blue points) of $A_D$ from the statistical noise are
 about ${\cal O}(10^{-3})$. Therefore, even a real anisotropic
 signal of ${\cal O}(10^{-3})$ exsists, it will be hidden
 behind the statistical noise.


\vspace{-11mm}  


\section{Concluding remarks}\label{sec4}

In the recent years, the field of FRBs is thriving and growing
 rapidly. It is of interest to study cosmology by using FRBs
 with known redshifts. In the present work, we try to test the
 possible cosmic anisotropy with the simulated FRBs. We find
 that at least 2800, 190, 100 FRBs are competent to find the
 cosmic anisotropy with a dipole amplitude 0.01, 0.03, 0.05,
 respectively. Unfortunately, even 10000 FRBs are not competent
 to find the tiny cosmic anisotropy with a dipole amplitude of
 ${\cal O}(10^{-3})$. On the other hand, at least 20 FRBs with
 known redshifts are competent to find the cosmic anisotropy
 with a dipole amplitude 0.1. We expect that such a big cosmic
 anisotropy can be ruled out by using only a few tens of FRBs
 with known redshifts in the near future.

Some remarks are in order. First, we find that so far FRBs are
 not competent to find the tiny cosmic anisotropy with a
 dipole amplitude of ${\cal O}(10^{-3})$. In fact, it is easy
 to imagine that even the cosmological principle is broken,
 the violation cannot be too large. For example, the Union2.1
 sample consisting of 580 SNIa suggests that there is a cosmic
 anisotropy with a dipole amplitude around
 $1.2\times 10^{-3}$~\cite{Lin:2016jqp}. No cosmic anisotropy
 has been found in the JLA sample consisting of
 740 SNIa~\cite{Deng:2018yhb,Lin:2015rza,Chang:2017bbi} and the
 latest Pantheon sample consisting of 1048 SNIa~\cite{Deng:2018jrp,
 Sun:2018cha,Andrade:2018eta}. Our results obtained here
 suggest that FRBs can be used to find or rule out the cosmic
 anisotropy with a dipole amplitude $\gtrsim 10^{-2}$, but
 cannot be used to find or rule out the cosmic anisotropy with
 a dipole amplitude $\lesssim {\cal O}(10^{-3})$.

Second, as mentioned in Sec.~\ref{sec3c}, a possible cosmic
 anisotropy of ${\cal O}(10^{-3})$ will be hidden behind the
 pseudo anisotropic signals of ${\cal O}(10^{-3})$ from the
 statistical noise. Thus, it is necessary to reduce the
 statistical noise. As mentioned in Sec.~\ref{sec2}, the main
 cause of this considerable statistical noise is the large
 $\sigma_{\rm IGM}\sim 100\;{\rm pc\hspace{0.24em} cm^{-3}}$
 \cite{McQuinn:2013tmc} due to the IGM plasma density fluctuations.
 On the other hand, actually this large $\sigma_{\rm IGM}\sim 100\;
 {\rm pc\hspace{0.24em} cm^{-3}}$ is also the cause of the relatively
 large uncertainty when one uses FRBs to constrain other cosmological
 parameters (see e.g.~\cite{Deng:2013aga,Yang:2016zbm,
 Gao:2014iva,Zhou:2014yta,Yu:2017beg,Yang:2017bls,Wei:2018cgd,
 Li:2017mek,Jaroszynski:2018vgh,Madhavacheril:2019buy,Wang:2018ydd,
 Walters:2017afr}). To obtain tight constraints, one has to
 combine FRBs with other cosmological probes such as SNIa, baryon acoustic
 oscillations (BAO), CMB, and GRBs. Therefore, it is of interest to reduce
 this large $\sigma_{\rm IGM}$ in FRBs cosmology.

Third, the present work can be extended to more general cases.
 For example, we consider a flat $\Lambda$CDM cosmology here,
 and it can be generalized to other cosmological models such as
 $w$CDM and CPL. On the other hand, here we only consider FRBs
 at redshift $z\leq 3$ to ensure that hydrogen and helium are
 both fully ionized. However, we can extend to high redshift
 $z\leq 6$. In the redshift range $3<z<6$, although helium is
 not fully ionized while hydrogen is fully ionized, the
 relevant calculations still can be carried out
 (see e.g.~\cite{Gao:2014iva}). Actually, it is expected that
 FRBs are detectable up to redshift $z\sim 15$ in
 e.g.~\cite{Zhang:2018fxt}. Although it is really a challenge
 to calculate DM at high redshift $z>6$, FRBs at high redshifts
 are fairly valuable in cosmology. Here, let us further discuss
 this issue in more details. If we consider a broader range of
 redshift e.g. $z\leq6$ for FRBs (we thank the anonymous
 referee~1 for pointing out this issue), helium is not fully ionized
 at $z>3$. In this case, $\chi_{e,\,\rm He}(z)$ becomes a piecewise
 function~\cite{Gao:2014iva}
 \be{eq26}
 \chi_{e,\,\rm He}(z)=\left\{
  \begin{array}{l}
  1\,,\quad\quad {\rm for}\quad z\leq 3\,,\\[2mm]
  0.025\,z^3-0.244\,z^2 + 0.513\,z + 1.006\,,
  \quad\quad {\rm for}\quad 3<z\leq 6\,.
  \end{array}\right.
 \ee
 If we extend to higher redshift $z>6$, hydrogen and helium are
 both not fully ionized, and hence $\chi_{e,\,\rm H}(z)$ and
 $\chi_{e,\,\rm He}(z)$ should be both piecewise functions,
 similar to Eq.~(\ref{eq26}). On the other hand, the fraction of baryon
 mass in the intergalactic medium $f_{\rm IGM}$ might be not a constant
 at high redshifts. In fact, the varying $f_{\rm IGM}(z)$ has
 been considered in the literature (e.g.~\cite{Li:2019klc,Wei:2019uhh,
 Qiang:2020vta}). In particular, it is reasonable to consider
 a linear parameterization with respect to the scale factor
 $a$, namely~\cite{Li:2019klc,Qiang:2020vta}
 \be{eq27}
 f_{\rm IGM}(z)=f_{\rm IGM,\,0}\,(1+\beta\,(1-a))=f_{\rm IGM,\,0}\,
 (1+\beta\,z/(1+z))\,.
 \ee
 So, if we consider FRBs at high redshifts $z>3$, both $f_e$
 and $f_{\rm IGM}$ might be functions of redshift $z$, and hence the
 calculation of $\langle{\rm DM_{IGM}}\rangle$ should be
 changed accordingly (nb. Eqs.~(\ref{eq12}) and (\ref{eq13})).

Fourth, we stress that our current results do not completely
 exclude the possibility that FRBs can become competent in the
 future to find or rule out the tiny cosmic anisotropy with a
 dipole amplitude $\lesssim {\cal O}(10^{-3})$. As mentioned in
 Sec.~\ref{sec1}, thousands FRB events per day over the entire
 sky are expected (see e.g.~\cite{Keane:2018jqo,
 Bhandari:2017qrj,Amiri:2019qbv}). So, numerous FRBs (say,
 $10^6$ FRBs, significantly more than 10000 FRBs considered
 in this work) might be available in the future. On the other
 hand, the statistical noise of FRBs (especially
 $\sigma_{\rm IGM}$) might be significantly reduced by the help
 of future developments (say, FRBs lensing and so on). And,
 many FRBs at high redshift $3<z<6$ might be also available in
 the future. Although FRBs with known redshift are very few
 by now, they will soon become numerous in the near future. In
 fact, very recently a non-repeating FRB 180924~\cite{Bannister2019}
 has been localized to a massive galaxy at redshift 0.3214 by using
 ASKAP. Another non-repeating FRB 190523~\cite{Ravi:2019alc}
 has been localized to a few-arcsecond region containing a
 single massive galaxy at redshift 0.66 by using DSA-10.
 Actually, several projects designed to detect and localize
 FRBs with arcsecond accuracy in real time are
 under constriction/proposition, for example,
 DSA-10~\cite{DSA-10} and DSA-2000~\cite{DSA-2000}. Thus, it is
 reasonable to expect that most of future FRBs are available
 with known redshifts. In summary, it is still possible that
 FRBs can become competent in the future to find or rule out
 the tiny cosmic anisotropy with a dipole
 amplitude $\lesssim {\cal O}(10^{-3})$, by the help of future
 developments mentioned above.

Fifth, in this work we assume that the redshift distribution of
 FRBs takes a form similar to the one of GRBs, namely Eq.~(\ref{eq18}).
 However, this might be changed as the number of FRBs increases in the
 future (we thank the anonymous referee~1 for pointing out this issue).
 In fact, there exist some different redshift distributions for FRBs in
 the literature. For example, two types of redshift distributions for
 FRBs have been proposed in~\cite{Munoz:2016tmg}, namely
 \be{eq28}
 P(z)\propto\frac{\chi^2(z)}{\left(1+z\right)H(z)}
 \,\exp\left(-\frac{d_L^{\,2}(z)}{2\,d_L^{\,2}(z_{\rm cut})}\right)\,,
 \quad{\rm or}\quad
 P(z)\propto
 \frac{\dot{\rho}_\ast(z)\,\chi^2(z)}{\,\left(1+z\right)H(z)}\,
 \exp\left(-\frac{d_L^{\,2}(z)}{2\,d_L^{\,2}(z_{\rm cut})}\right)\,,
 \ee
 where $\chi(z)=d_L(z)/(1+z)=c\int_0^z d\tilde{z}/H(\tilde{z})$ is the
 comoving distance, and $d_L(z)$ is the luminosity distance.
 $\dot{\rho}_\ast(z)$ corresponds to the star-formation history (SFH).
 Gaussian cutoff at $z_{\rm cut}$ is introduced to represent an
 instrumental signal-to-noise threshold. Of course, other types
 of redshift distributions for FRBs are also possible. However, we argue
 that different redshift distributions of FRBs do not significantly change
 the results obtained in this work. As shown in
 Eqs.~(\ref{eq20})--(\ref{eq25}), the dipole anisotropy is mainly related
 to the positions of FRBs in the sky (namely right ascension $\alpha$
 and declination $\delta$), rather than their redshifts. Thus, it is
 reasonable to expect that the redshift distribution of FRBs will not
 play an important role in testing the cosmic anisotropy.

Sixth, as mentioned in the beginning of Sec.~\ref{sec2}, we use $\rm DM_E$
 instead of $\rm DM_{obs}$ to study the FRB cosmology for convenience
 following e.g.~\cite{Yang:2016zbm,Yang:2017bls}. Actually, the difference
 between them is the contribution from Milky Way, $\rm DM_{MW}$. For a
 well-localized FRB, the corresponding $\rm DM_{MW}$ can be known by
 using the NE2001 model~\cite{Cordes:2003ik,Cordes:2002wz} or the YMW16
 model~\cite{YMW16} for the galactic distribution of free electrons and
 its fluctuations constructed from the known pulsar DM data. It
 is worth noting that we have not neglected $\rm DM_{MW}$ in fact (although
 $\rm DM_{MW}$ is usually small with respect to $\rm DM_{obs}$). Instead,
 we just subtract $\rm DM_{MW}$ from $\rm DM_{obs}$ to introduce $\rm DM_E
 =DM_{obs}-DM_{MW}$, as in Eq.~(\ref{eq6}). Since there is no theoretical
 model for $\rm DM_{obs}$, we instead calculate $\rm DM_E\equiv DM_{obs}
 -DM_{MW}=DM_{IGM}+DM_{HG}$ in the simulations because $\rm DM_{obs}=DM_{MW}
 +DM_{IGM}+DM_{HG}$ (see Eq.~(\ref{eq5})). Accordingly, the uncertainty of
 $\rm DM_E$ is given by $\sigma_{\rm E}=(\sigma_{\rm IGM}^2+
 \sigma_{\rm HG}^2)^{1/2}=\{\,\sigma_{\rm IGM}^2+[\,\sigma_{\rm HG,\,loc}/
 (1+z)\,]^2\,\}^{1/2}$ in the simulations, as mentioned at the end of
 Sec.~\ref{sec3a}. Of course, one can also calculate $\sigma_{\rm E}$ by
 using the uncertainties of $\rm DM_{obs}$ and $\rm DM_{MW}$, since
 $\rm DM_E=DM_{obs}-DM_{MW}$. Although the uncertainty of $\rm DM_{obs}$
 is usually small, $\sigma_{\rm E}$ is not so small because we have not
 neglected $\rm DM_{MW}$ and hence the uncertainty of $\rm DM_{MW}$
 should be also taken into account. Since $\rm DM_{MW}$ is known by
 using the NE2001 model~\cite{Cordes:2003ik,Cordes:2002wz} or the YMW16
 model~\cite{YMW16} for the galactic distribution of free electrons
 and its fluctuations, it is model-dependent in some sense, while its
 uncertainty $\sigma_{\rm MW}$ is not small in fact (on the other hand, one
 should be aware of the considerable difference between the NE2001 and
 YMW16 models). Because we are mainly concerned about the cosmological
 information carried by $\rm DM_{IGM}$, while it is hard to construct
 $\rm DM_{MW}$ from the NE2001 or YMW16 models and there is no theoretical
 model for $\rm DM_{obs}$, in this work we choose to simulate $\rm DM_E$ by
 using $\rm DM_{IGM}$ and $\rm DM_{HG}$, as well as their uncertainties.

Seventh, we have paid attention to the possible dipole with amplitude of
 ${\cal O}(10^{-3})$ or ${\cal O}(10^{-2})$ in this work (we thank the
 anonymous referee~2 for pointing out this issue). In the literature, the
 claimed cosmic dipole amplitude is $A_D\sim{\cal O}(10^{-3})$
 or ${\cal O}(10^{-2})$ by using SNIa (see e.g.~\cite{Mariano:2012wx,
 Cai:2011xs,Yang:2013gea,Chang:2014nca}), SNIa+GRBs (see
 e.g.~\cite{Wang:2014vqa,Chang:2014jza}), radio galaxies (see
 e.g.~\cite{Bengaly:2017slg}), and so on. As is well known, the dipole
 in the CMB temperature map is also $A_D\sim{\cal O}(10^{-3})$ (see
 e.g.~\cite{Aghanim:2013suk}). These results in the literature motivate
 us to consider the possible dipole in FRBs with amplitude of similar
 order of magnitude.

Eighth, we would like to emphasize the physical significance of a possible
 cosmic dipole (we thank the anonymous referee~2 for pointing out this
 issue). Here we are mainly interested in probing differential expansion of
 space, as in e.g.~\cite{Antoniou:2010gw,Cai:2011xs}. A cosmic dipole gives
 a hint of one (or more) preferred axis in the universe. Actually, the
 preferred axis is possible in various exotic cosmological models. For
 example, in the well-known G\"odel solution~\cite{Godel:1949ga} (see also
 e.g.~\cite{Li:2016nnn} and references therein) of Einstein's field
 equations, the universe is rotating around an axis. As is well known, the
 time travel is possible in the G\"odel universe. Besides the G\"odel
 universe, in the Finsler universe~\cite{Li:2015uda} and most of Bianchi
 type I$\,\sim\,$IX universes~\cite{Bianchi}, the cosmic anisotropy is also
 possible. So, if a cosmic dipole is confirmed, it will force us to
 consider such kind of exotic models seriously.

Ninth, it is of interest to discuss the potential astrophysical
 systematics contaminating the signal. In particular, one might speculate
 that observational systematic errors could create an apparent dipole
 signal for FRBs, for example, if more high redshift FRBs are observable in
 one part of the sky as compared to another due to foreground contamination
 (we thank the anonymous referee~2 for pointing out this issue). Here we
 argue that such kind of systematics might be fairly minor. It is useful to
 consult with the cases of SNIa. The Pantheon sample~\cite{Scolnic:2017caz,
 Pantheondata,Pantheonplugin} is the largest spectroscopically confirmed
 SNIa sample to date, which consists of 1048 SNIa. As is shown in Fig.~1
 of~\cite{Deng:2018jrp} (and its Sec.~2), more than one half of 1048
 Pantheon SNIa are located in a small region of the southern sky (because of
 some unknown reasons to our knowledge). However, no evidence for the cosmic
 dipole is found in the Pantheon SNIa sample~\cite{Deng:2018jrp}. In fact,
 it is the same for the JLA sample consisting of 740 SNIa~\cite{Lin:2015rza,
 Chang:2017bbi}. In light of the cases of SNIa, we expect
 that the potential systematics mentioned above might be fairly minor.

Finally, the field of FRBs is growing rapidly. In fact, many
 new findings have been obtained after the (pre-)commissions
 of ASKAP and CHIME in 2018. Big breakthroughs in the coming
 years are expected. Therefore, FRBs cosmology might also have
 a promising future.


\section*{ACKNOWLEDGEMENTS}

We thank the anonymous referees~1 and 2 for quite useful comments and
 suggestions, which helped us to improve this work. We are grateful to the
 anonymous referee~3 (CQG Advisory panel member) for fair judgement. We
 also thank Profs. Xue-Feng~Wu, Zheng-Xiang~Li, He~Gao, Yuan-Pei~Yang,
 Bin~Hu, Lixin~Xu, Puxun~Wu, Xin~Zhang, Qing-Guo~Huang, Zhoujian~Cao,
 Jun-Qing~Xia, and Zong-Hong Zhu for useful discussions on FRBs during
 the period of Workshop on Multi-messenger Cosmology, BNU,
 $9\sim 11$ November 2018. We thank Zhao-Yu~Yin, Xiao-Bo~Zou,
 Zhong-Xi~Yu, Shou-Long~Li and Shu-Ling~Li for kind help and
 discussions. This work was supported in part by
 NSFC under Grants No.~11975046 and No.~11575022.

\renewcommand{\baselinestretch}{1.01}


\end{document}